\documentstyle[emulateapj,epsfig,apjfonts]{article}

\def \etal { et~al.\ }  
\def\nuc#1#2{\relax\ifmmode{}^{#1}{\protect\text{#2}}\else${}^{#1}$#2\fi}

\def\msol#1{\relax$#1\,M_\odot\/$}

\def\pagefig#1#2#3#4{
\begin{figure*}[t]
\plotfiddle{#1}{#2}{0}{67}{67}{-210}{0}
\caption{\footnotesize
#3
\label{#4}}
\end{figure*}
}

\def\colfig#1#2#3#4{
\vskip 0.5cm
\refstepcounter{figure}
\epsfig{file=#1,width=#2}
\noindent{
\footnotesize
F{\scriptsize IG}.~~\thefigure.---
#3
\vskip 0.5cm
}
\label{#4}
}

\begin{document}

\title{Further Adventures: Oxygen Burning in a Convective Shell}
\author{S.M. Asida\altaffilmark{1}
and David Arnett\altaffilmark{2}}
\affil{Steward Observatory, University of Arizona ,Tucson, AZ 85721, USA}
\authoremail{sasida@as.arizona.edu}
\altaffiltext{1}{E-mail: sasida@as.arizona.edu}
\altaffiltext{2}{E-mail: darnett@as.arizona.edu}
\begin{abstract}
Two dimensional hydrodynamical simulations of convective oxygen burning shell in 
the presupernova evolution of a \msol{20} star are extended to later times. We 
used the  VULCAN code to simulate longer evolution times than previously 
possible. Our results confirm the previous work of \cite{ba98} over their time 
span (400~s). However, at 1200~s, we could identify a new steady state that is 
significantly different than the original one dimensional model. There 
is considerable overshooting at both the top and bottom boundaries of convection
zone. Beyond the boundaries, the convective velocity falls off exponentially, 
with excitation of internal modes.  The resulting mixing greatly affect the 
evolution of the simulations. Connections with other works of simulation of
convection, in which such behavior is found in a different context, are 
discussed.

\end{abstract}


\section{Introduction}

There are many published studies of stellar convection using multidimensional
hydrodynamical simulations, but few deal with convective nuclear
burning occurring in stellar interior (see \cite{dpr98} for a study of core 
hydrogen convection). \cite{arn94} (A94) and \cite{ba94,ba98}
(BA94, BA98) have studied oxygen burning shell which is one of the last stages 
in a massive star presupernova evolution; we extend that work.

This is an important stage in presupernova evolution because in this 
convective region several phenomena take place. In or near this region:
(1) most of the explosive nucleosynthesis and production of 
\nuc{56}{Ni} and \nuc{57}{Ni} occurs, (2) the ``mass cut'' between collapsed
and ejected matter develops, and (3) mixing of different layers may happen. 
The standard model for treating convection in one dimensional (1D) stellar 
evolutionary codes, the mixing length theory (MLT), is usually used for 
modeling this convective region as well, even 
though the conditions of oxygen convective 
burning shell are more complicated than can be assumed for MLT to be valid 
(i.e., the flow is not strongly subsonic and there are 
both energy sources and sinks in the flow).

In \cite{arn94} this evolutionary stage was described, and the first two 
dimensional (2D) hydrodynamical simulation of this problem were presented. 
The 145~s time interval was simulated by these calculations, using the 
PROMETHEUS code, is much less than the duration of this stage 
($10^3$ to $10^5\rm\ s$, see figures 10.5 and 10.6 in \cite{arn96}). 
During this time, convective flow was formed. \cite{ba94,ba98} have 
modeled the evolution over a longer time interval of 300 to 400~s, 
and noticed a mixing of composition 
from the neighboring stable region near the end of the simulations. 
In addition,
\begin{itemize}
\item the flow velocities were up to $10\%-20\%$ of the sound speed,
\item stellar structure was not altered significantly, and 
\item there were strong density and composition fluctuations 
in both space and time which did not reach a statistical steady state. 
\end{itemize}

The main differences between this work and that of A94, BA94, and BA98
 were (1) the use of a different hydrodynamic code, and (2) the 
simulation of a longer evolution time.  The same initial model, 
equation of state, and nuclear reaction algorithms were used.

\pagefig{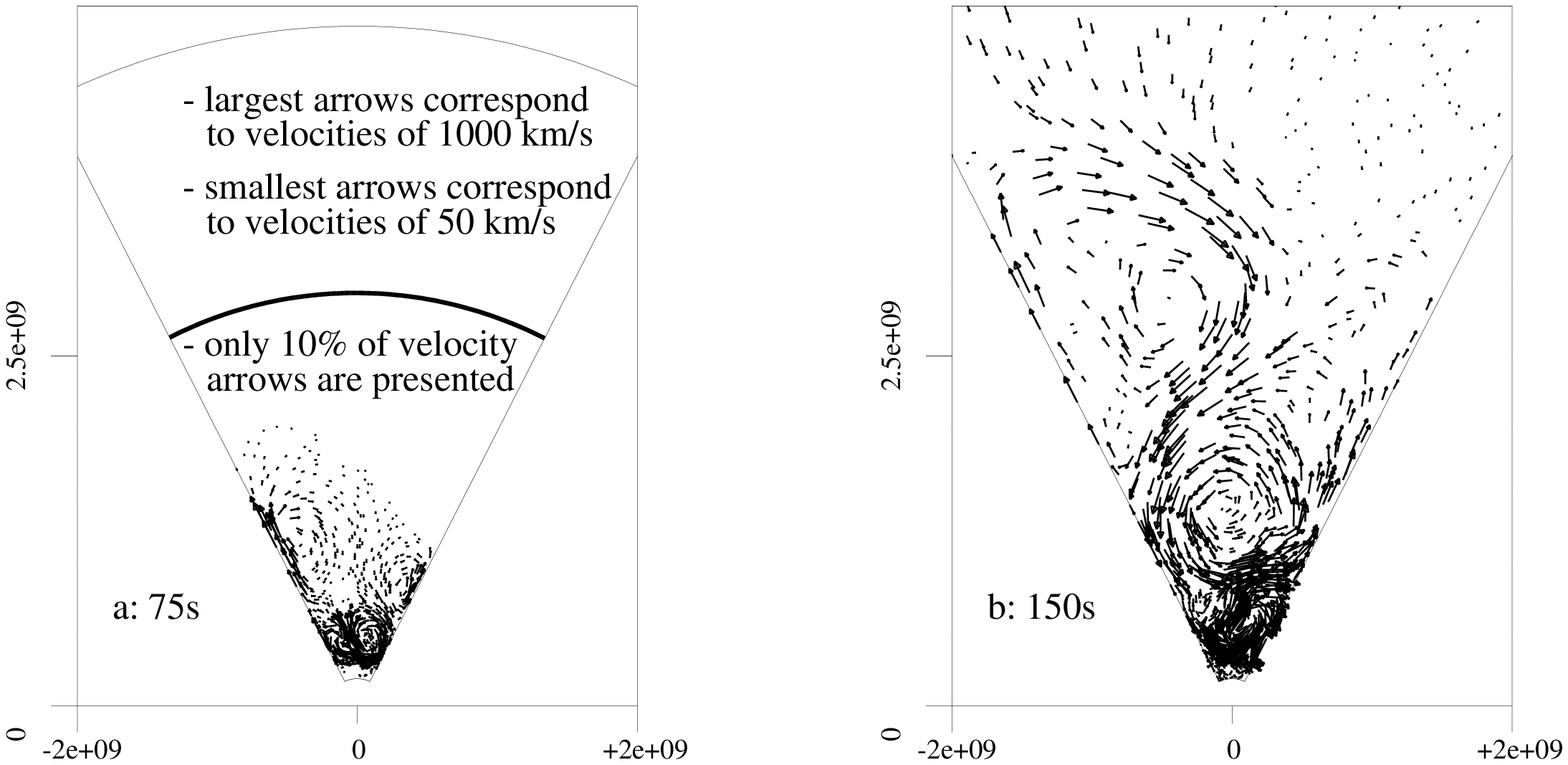}{9cm}{
Two dimensional velocity field for: a - 75~s, b - 150~s. The upper boundary
of the oxygen shell is presented by the thick solid line in a.}{f1}

\section{The numerical scheme}

For this study, we used a version of the hydrodynamical code VULCAN 
(\cite{lvn93}). This code uses an algorithm that begins with a  hydrodynamic 
time step, and is completed by a relaxation of the numerical grid, thus giving 
an Arbitrary Lagrangian Eulerian (ALE) scheme. In the hydrodynamic time step, 
the Lagrangian hydrodynamic equations in two dimensions are solved explicitly 
or implicitly, allowing longer time steps to be taken if the flow is strongly 
subsonic. The mesh relaxation phase is necessary to eliminate distortion of the 
cells, especially for flow with vorticity. We used a quasi-1D Lagrangian 
relaxation, in which each radial row of cells kept a constant mass. The code 
was used by \cite{gl95} for computing convective novae outbursts, and by 
\cite{at97} and \cite{asd00} to simulate convection in a red giant envelope. 
One of the adaptation made by \cite{asd00} for simularions of convection was 
the inclusion of \cite{smg63} like sub grid scale mixing (SGSM) model, as was
done in many two dimensional and three dimensional numerical studies of 
convection.

The equation of state and the thermonuclear reactions algorithms were 
essentially the same as in A94 and BA98. The equation of state was the sum 
of components for electrons, ions, and radiation.
The nuclear reaction network used twelve species for  helium, carbon, 
neon and oxygen burning.  Neutrino cooling was included; see the above 
references for details. The initial model was the same as in A94 and BA98.
The computational domain corresponds to a region containing the entire
oxygen-burning shell in a star of \msol{20}, having an 
initial metallicity of 0.007 (about one third solar). This shell boundaries 
are at radii of $3\times 10^8\rm cm$ and $3\times 10^9\rm cm$ that are 
equivalent to mass coordinates of \msol{1.4} and \msol{2.4}.

We performed several simulations that were different in: the computational 
domain, spacial resolution, the initial temperature gradient and numerical
parameters of the simulation. The computational domain typically includes 
all of the oxygen shell as well as the neighboring layers (the inner radius
was located at $2\times 10^8\rm cm$ in some of the simulations, and the outer
radius was located at $8\times 10^{10}\rm cm$ in other simulations). The 
angular extent of the wedge ran usually from 0.35 $\pi$ to 0.65 $\pi$ radians.

Typically. the oxygent shell was divided to $\sim $ 120 radial zones, and the 
spacing of zones was logarithmic in radius and linear in angular direction 
(i.e., $dr = r\ d\theta $) with 60 angular zones. Rotational symmetry was 
assumed. These characteristics are similar to the medium resolution 
simulations of BA98. Because of small differences in the equation of state 
in the initial one dimensional model and in the two dimensional code, as 
well as different radial zoning, we had two sets of simulations: in the first,
the 1D model (with interpolation) was used as it is, and in the second it was
slightly adapted, so that the temperature gradient would be superadiabatic
in all zones of the convection shell.

The velocities on the inner boundary were set to be zero for the whole 
simulation, so that the inner core was a hard sphere. At the upper boundary 
there was no limitation on the velocities, but in order to eliminate mass 
flow out of the computational domain, an average radius was used to follow 
expansion or shrinking of the outer boundary. 
We used reflective boundary conditions on the sides (though such conditions 
enforce a downflow or upflow on the side boundaries, it was found in BA98
not to be important).

\section{Results of the simulations}

\pagefig{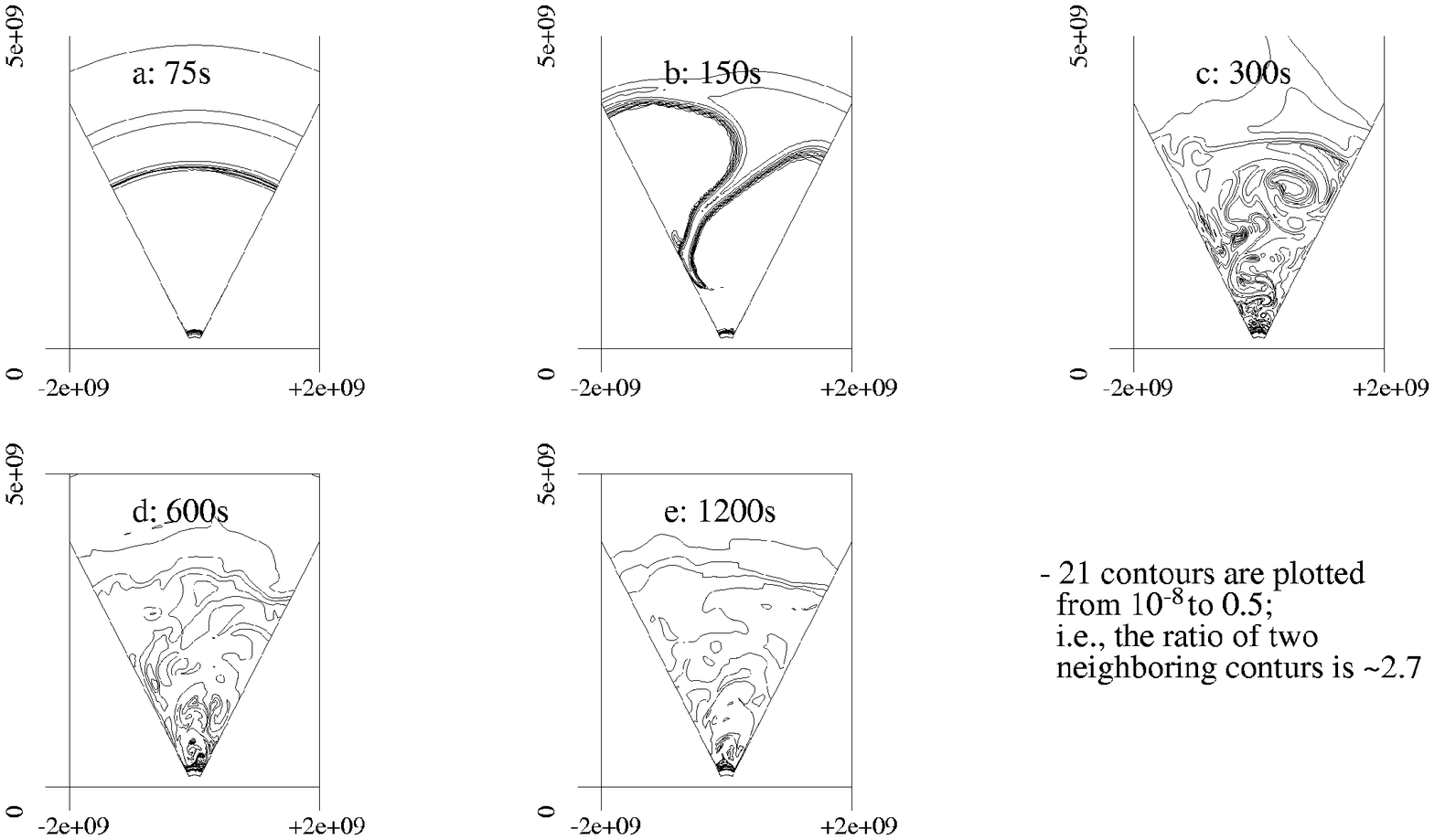}{9cm}{
Carbon abundance contours for: a - 75~s, b - 150~s,
c - 300~s, d - 600~s, e - 1200~s.}{f2}

We present the results of one ``standard'' simulation with 172 radial zones
and 60 angular zones. The inner boundary was at $2\times 10^8\rm cm$ and the
outer was at $48\times 10^8\rm cm$. The temperature gradient was slightly super
adiabatic in the oxygen shell, and no SGSM terms were used. Most of the results 
of the other simulations were similar and the differences are discussed mainly 
in the end of this section.

\subsection{General Evolution of the Flow}

In our simulations, the initial velocities were zero, and the convective flow 
developed as a result of the instability from round off errors. Figure 1 
presents the velocity field in the beginning of the simulations for times (a) 
75~s and 150~s. As we can see, the convective flow starts at the bottom of the 
convection region (i.e., the burning layer), and then moves up with increasing 
eddy size. By time 150 seconds the convective flow penetrates the upper boundary 
of the convection region (seen as a thick line in panel a). As a result, there 
is a downflow of carbon-rich material. This can be seen in Figure 2, which 
presents contours of carbon nucleon fraction. Panels~a-e represent times of 75, 
150, 300, 600 and 1200~s. We can see that the downflow (panel b) penetrates the 
whole convective region, and results in mixing of carbon in this region. From 
comparison of panels c, d and e we can see that carbon abundance becomes more 
uniform, as the simulation evolves.

This penetration of carbon is almost identical to that seen by 
BA98; see their Figure 5. The two independent hydrocodes give consistent 
results over the whole time spanned (400 seconds) by BA98 simulations. The 
small difference in the timescale for the carbon penetration is due to the 
small differences in the extent of the super 
adiabatic gradient of the initial 1D model (when we used the initial 1D model
as it is, without modifying the temperature gradient, we got very similar 
results with a longer timescale of the penetration).

\newpage
\subsection{Carbon Enrichment}

Carbon penetration, as well as other processes in the simulations, can be 
visualized by examining one dimensional averages of various parameters, which 
in principle would be equivalent to results from 1D simulations. In Figure 3 
carbon abundance (nucleon fraction) is plotted as a function of stellar mass 
coordinate. From the initial profile (solid line) we can see the location of 
the oxygen shell. At 150~s (long dashed line), carbon rich material has entered 
the convection layer, and then mixed through the whole region. At later times, 
the fluctuations in the abundances of carbon decrease (compare the profiles at 
300~s -short dashed line, 600~s - dotted line and 1200~s -dotted dashed line), 
and we have a carbon nucleon fraction of $\sim 5\times10^{-3} $. 

\colfig{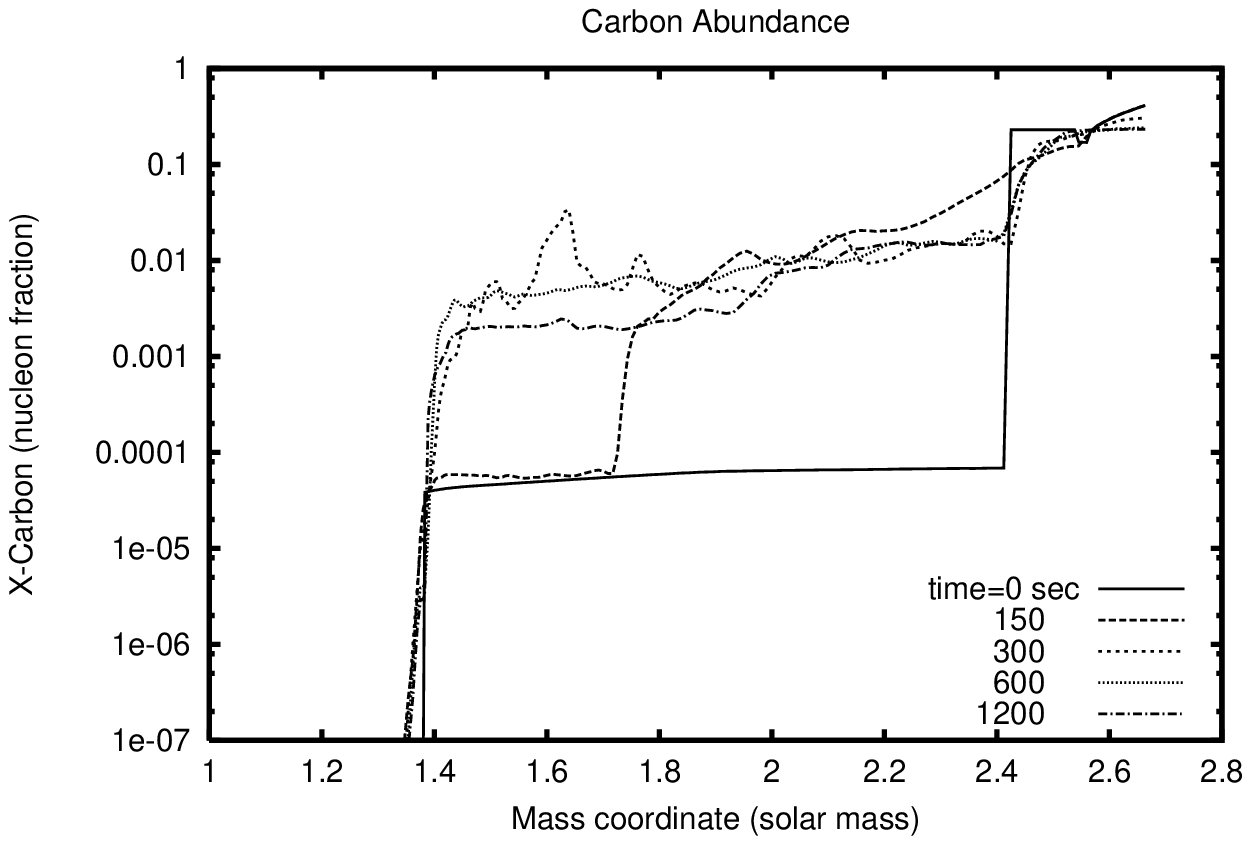}{8.0cm}{Average carbon abundance}{f3}

A change in abundance in this plot does {\it not} necessarily corresponds to
mixing, because this plot represents an one dimensional average of the 
compositions of all the material in each radial layer. This may be revealed by 
a closer look at the apparent widening of the jump in carbon abundance at the 
upper boundary of the oxygen shell at $m \approx $\msol{2.4}: this jump is much 
wider at 150~s than it is at later times. This ``anti diffusion'' is possible 
since the widening of the interface of the shell is the outcome of two 
processes: a mixture of material from the two sides of the interface {\it and} 
large scale motions that change the shape of the interface. Thus, at later times
 the interface is more spherical than it was at 150~s, so that averaging the 
abundances over radial layers yields a sharper jump (compare Fig 2 panels b and
 d).

\colfig{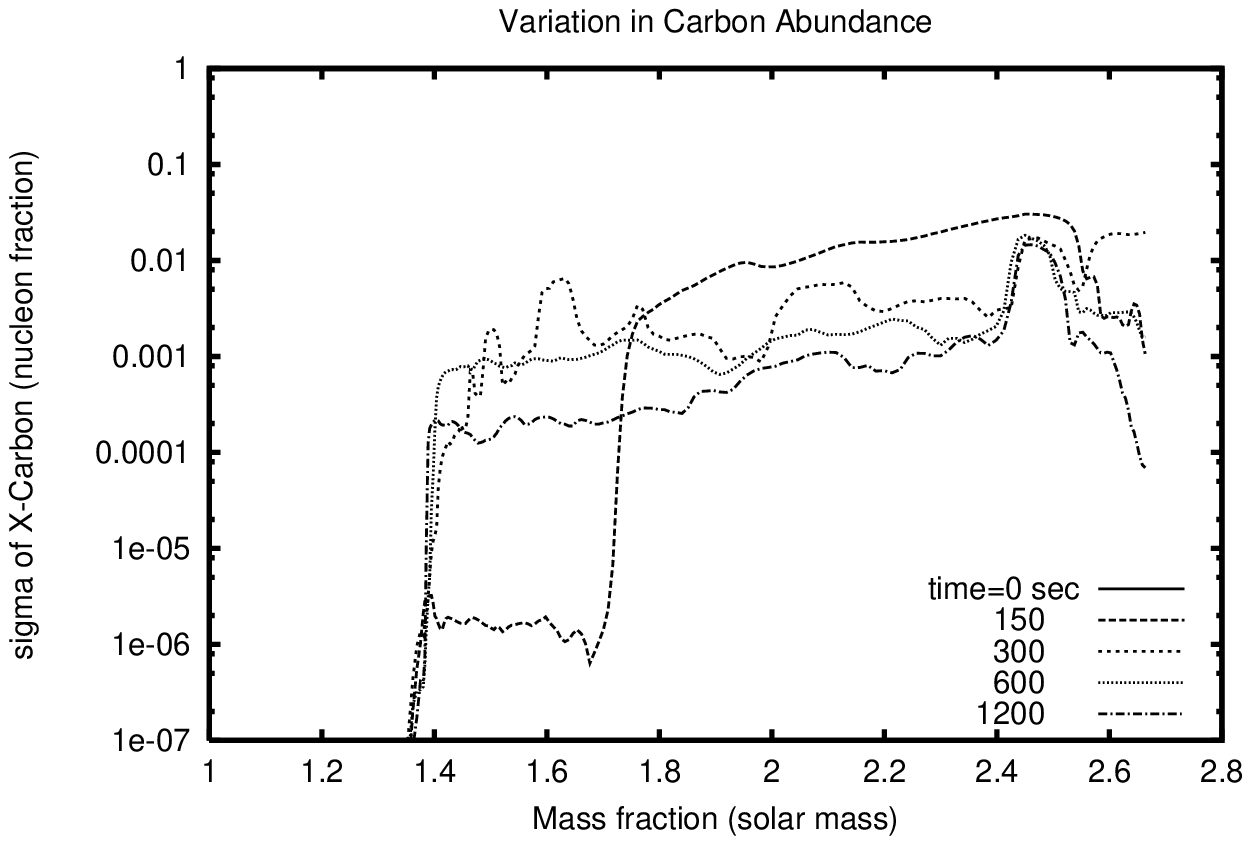}{8.0cm}{RMS fluctuations of carbon abundance}{f4}

To better understand the carbon enrichment we present Figure 4 which 
shows the RMS fluctuations of carbon nucleon fraction ($\sigma$) as a function 
of stellar mass coordinate at several different times. The tendency toward
more uniform carbon mixing is clearly seen, as was indicated by Figures 
3 and 4. From this figure, we can also see, that the previously
mentioned widening of the oxygen shell interface is partially due to changes
in its shape since the fluctuations in carbon fraction at the boundary are
relatively high, and are even higher at time 150~s.

\colfig{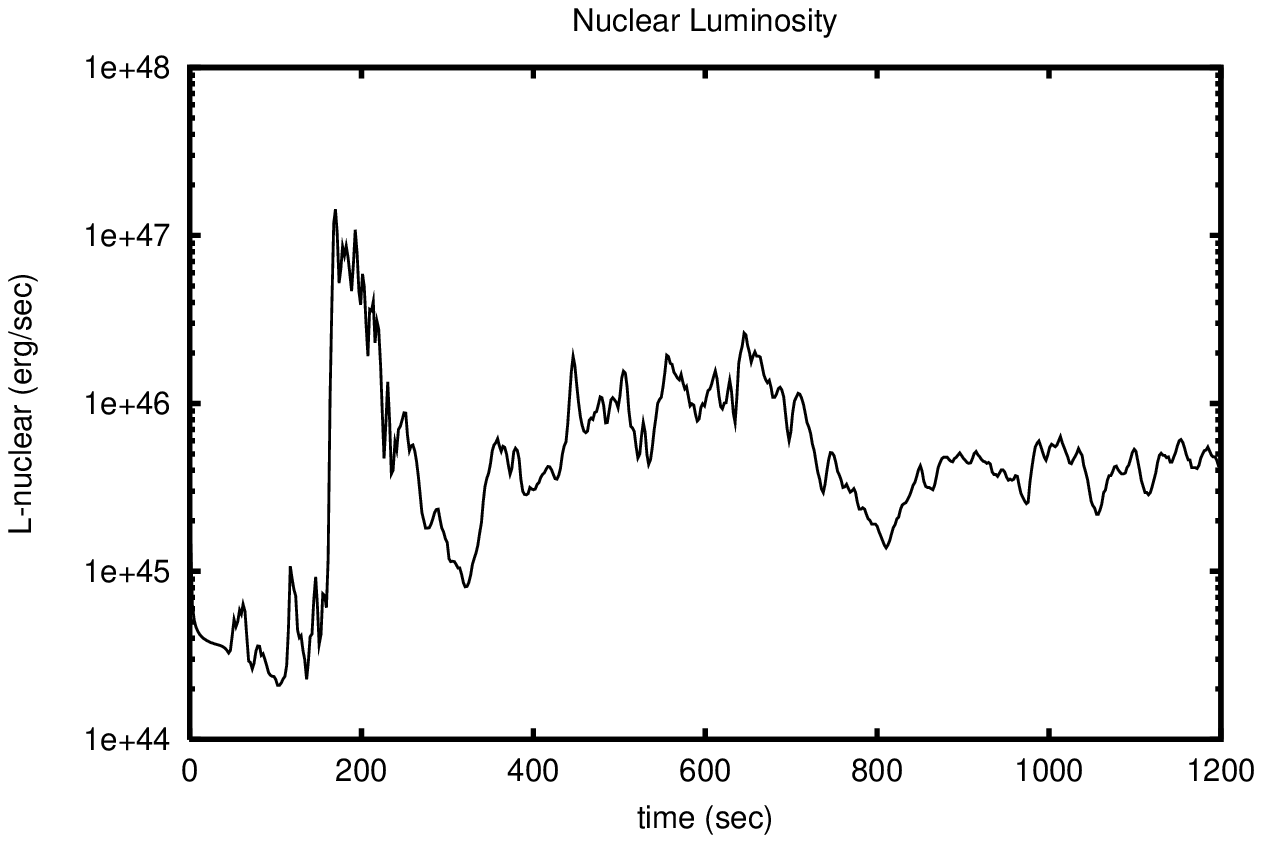}{8.0cm}{Nuclear luminosity as a function of time.}{f5}

The mixing of carbon in the burning layer causes the reaction rate to change 
significantly. In Figure 5, the nuclear luminosity is presented as a 
function of time. In the beginning there is an adjustment phase in which the 
energy production rate decreases to about $5\times 10^{44}\rm\ erg\ s^{-1}$ .
 This transient phase is a thermal relaxation due to the fact that the initial 
model is inconsistent: it does not have a two dimensional velocity field which 
can carry the convective energy flux that the 1D model needed. As 
the carbon rich material penetrates to high temperature layers, it starts to 
react rapidly, yielding nuclear luminosities which increase to about hundred 
times the initial value. In this stage, the result is mainly carbon and neon 
consumption. Afterward, there is a smaller decrease in nuclear luminosity
(compare time 600 and 1200~s). This
decrease is related to the small decrease in the average carbon abundance seen
between times 600~s and 1200~s in  Figure 3.

\subsection{A Quasi-Steady State}

\colfig{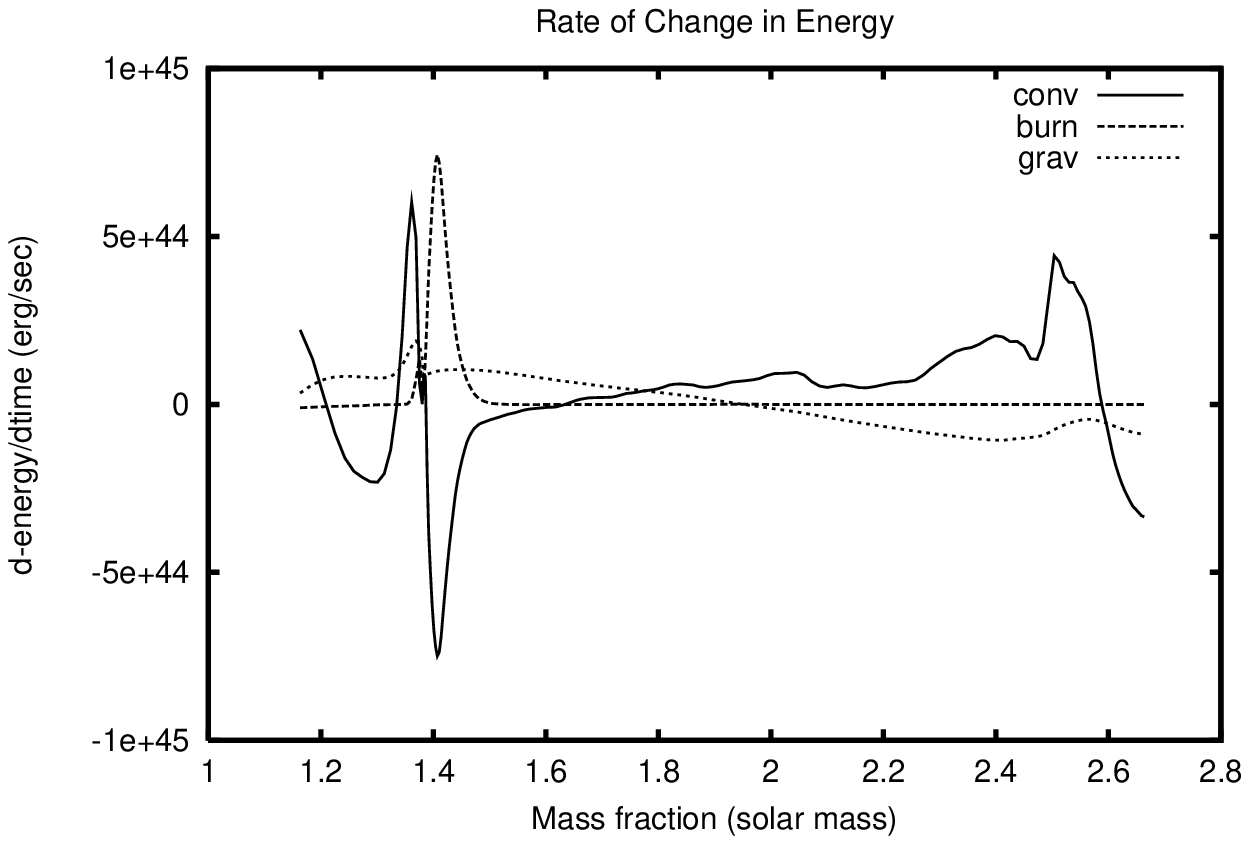}{8.0cm}{Time average change in energy}{f6}

The relaxation in the carbon abundances by time 300~s to a slowly varying 
state, and the small change between times 600 and 1200~s, suggest that the 
system may be close to a quasi - stationary state (a ``steady'' state on 
average). This one is significantly different from the steady state predicted 
by the one dimensional calculations used to definethe initial model. That the 
system is close to a steady state can be demonstrated by the following figures.
Figure 6 shows the average rate of change in energy, as a function of 
mass coordinate (from time 400 to time 800~s). The energy produced by 
thermonuclear reactions (dashed line) is carried away by convection 
(solid line). The net change in energy is small, and balanced by changes 
in the gravitational energy (dotted line). This slow change is an adjustment 
of the structure driven by enhanced nuclear burning from the carbon ingestion.
There is a net heating at the bottom, below \msol{1.4}, which will be addressed
in the next section.

\colfig{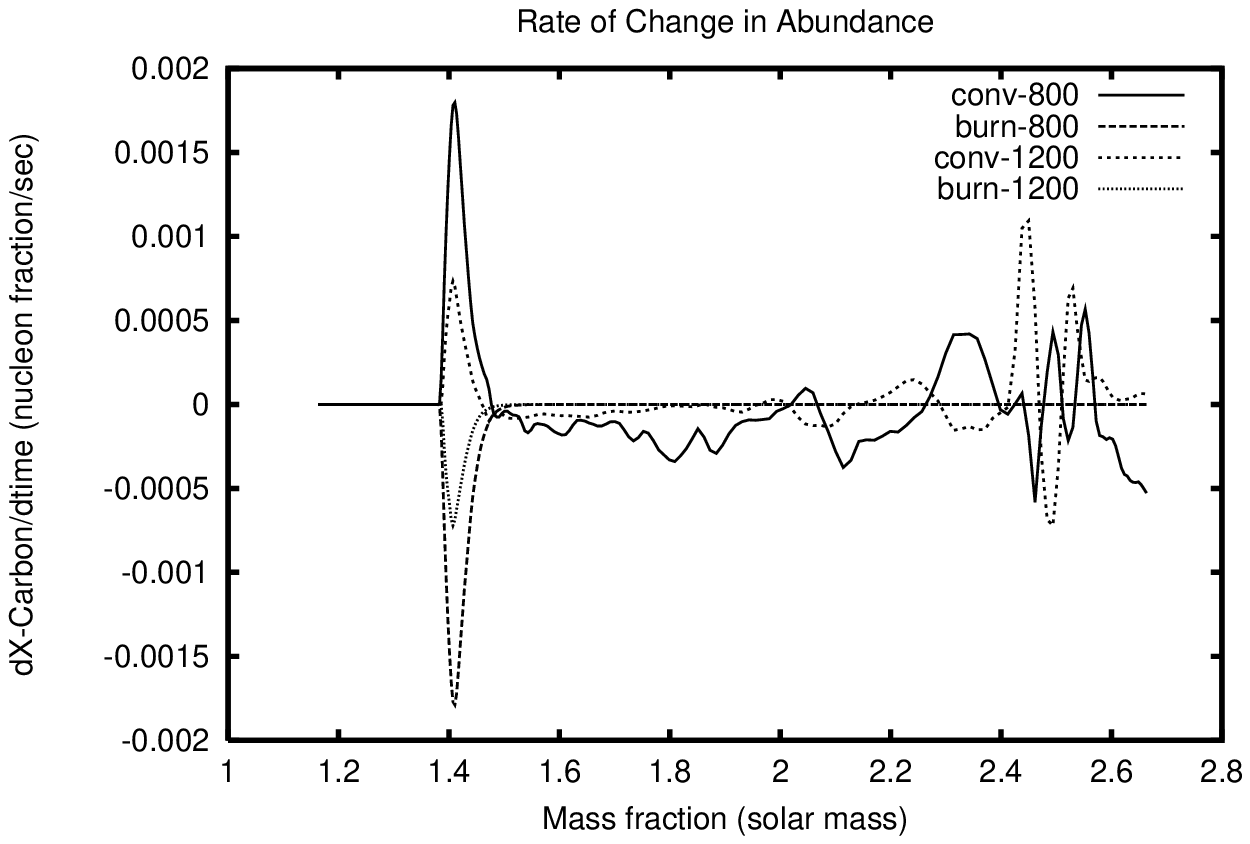}{8.0cm}{Time average change in carbon abundance}{f7}
 
Figure 7 shows the average rate of change in carbon nucleon fraction, 
as a function of mass coordinate (from time 400~s to time 800~s, and from time 
800~s to time 1200~s), comparing the changes due to the divergence of the 
convective abundance flux (solid line) with that due to nuclear burning (long
dash line). The nuclear consumption of carbon at the bottom is almost completely 
balanced by convective inflow; This is the burning zone. Over most of the 
convection region the change in carbon abundances is small. There are larger 
fluctuations at the top (outer) boundary of the convection zone (where the 
carbon abundance increases by a factor of $\sim 20$, see Fig. 3).

The decrease in nuclear luminosity (seen in Figure 5, after the peak at 200~s) 
was caused by a decrease in the net flux of carbon into the burning layer. This 
is easily seen in Figure 7 in the comparison of the changes in the later time 
interval (800~s to 1200~s - short dashed line) to the changes in the previous 
time interval (400~s to 800~s - solid line). The total mass of carbon in the 
oxygen convection shell is decreasing as a result of carbon consumption and 
less mixing from the upper shells. Thus our convective burning is beginning to 
evolve on a secular time scale by the consumption of fuel, having approached a 
thermal ``steady'' state.

in Figure 8 we present the RMS fluctuations of density (divided by the density)
 as a function of stellar mass coordinate for the standard simulation for 
several times. As the composition becomes more uniform with time, the density
fluctuation decreases throughout most of the oxygen shell to about $10^{-3}$ at 
time 1200~s. However at the edges of the convection zone, the fluctuations does
not decrease and we have about $2\times 10^{-2}$ at the bottom and 
$7 \times 10^{-2}$ at the top, where there is a jump in the abundances. 
These values are similar to those obtained by BA98.

\colfig{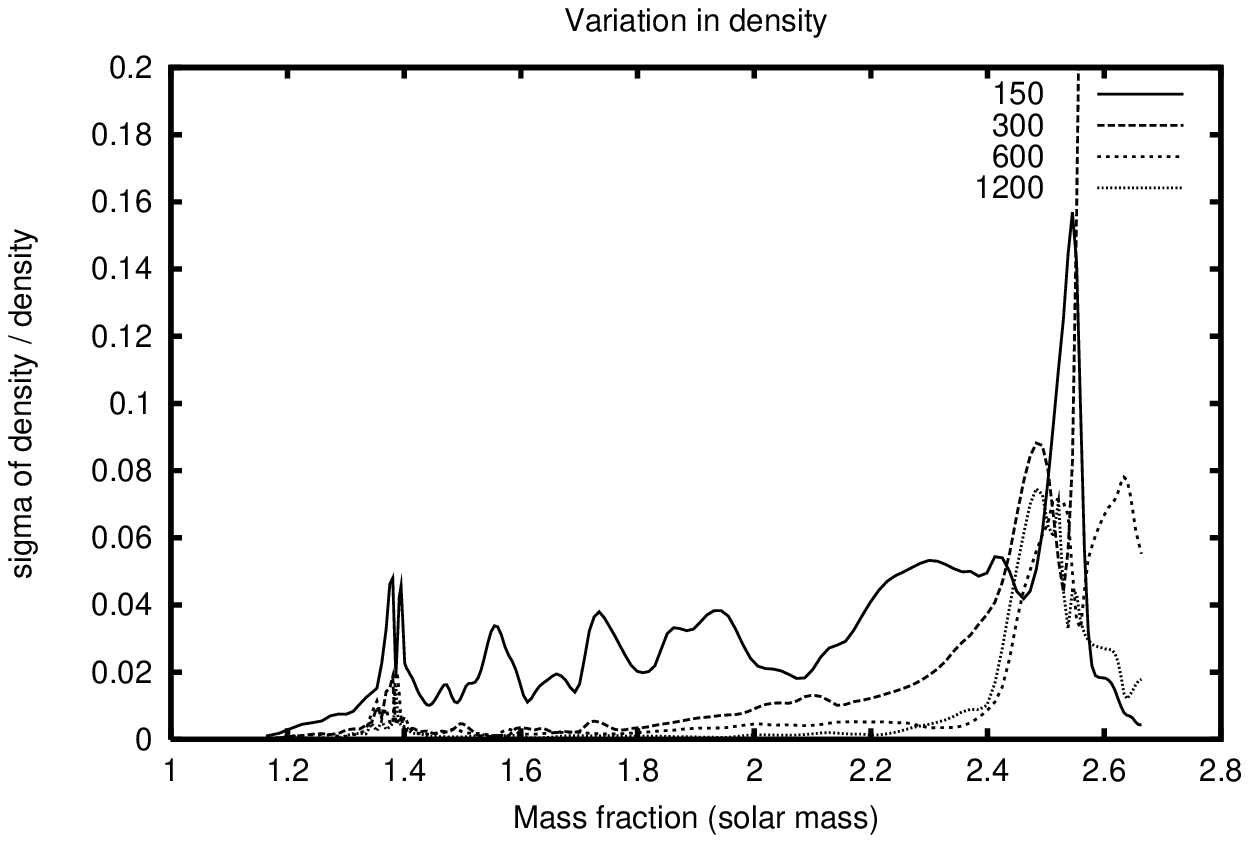}{8.0cm}{RMS fluctuations of the density}{f8}

\subsection{Interaction with Neighboring Shells}

\colfig{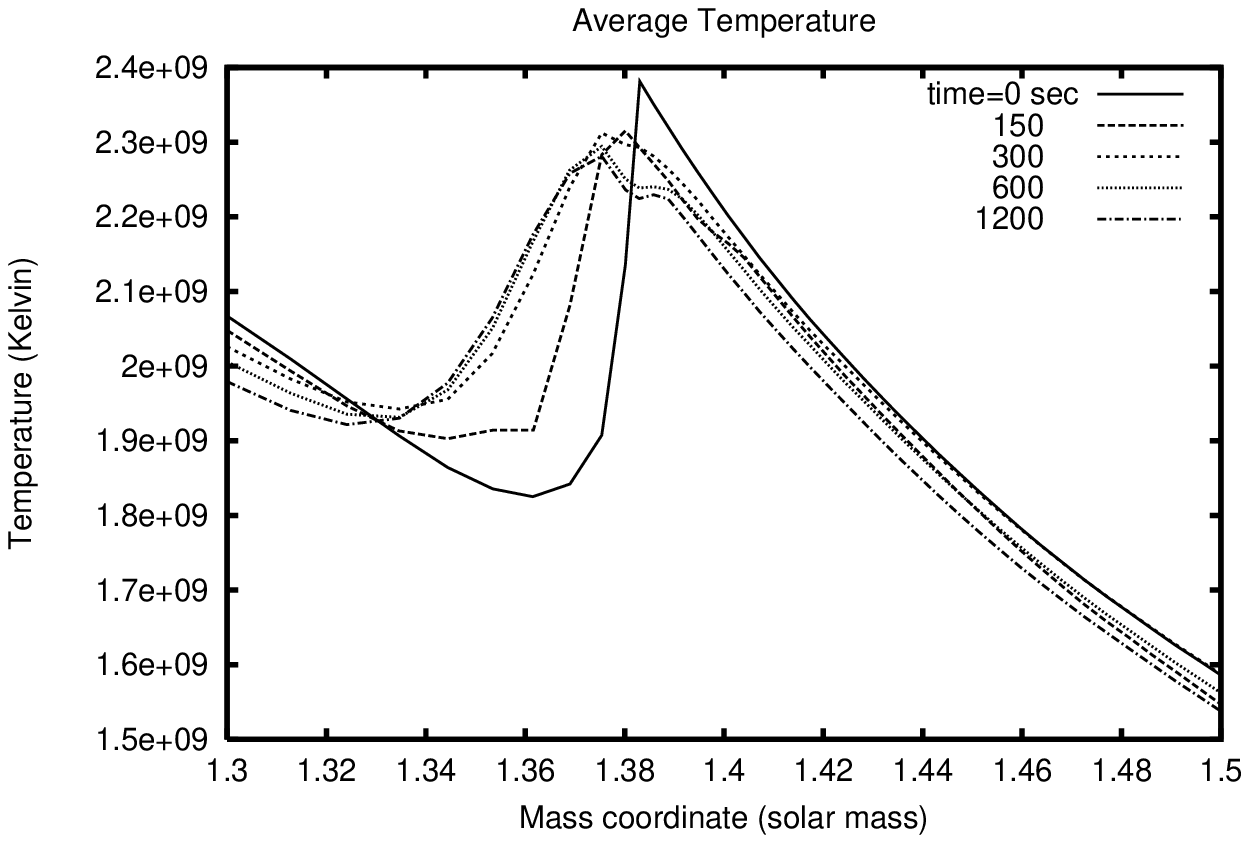}{8.0cm}{Temperature profile at the bottom}{f9}

One important interaction of the convective shell with neighboring shells is 
the mixing of carbon from the upper carbon rich shells, as described before. 
Another important interaction is heating of the nearest shells below the bottom
of the convection zone: in the initial model there is a sharp decrease in 
temperature just below the convection zone, as convective flow starts to grow
inside the lower parts of the convection zone there is some penetration of the
flow to these lower temperature shells. As a result of this flow, the lower
temperature shells are being heated and the higher temperature shells are 
being cooled. This heating is easily seen in Figure 9 where we present the 
temperature profile in the initial model, and at later times. Most of the 
heating was done by time 300~s, but even at later times, some heating exsisted,
as can be seen in the left end of Figure 6. Along with this ``energy mixing'' 
there is mixing of composition in those few zones as we can see from neutron 
excess $\eta$ plot presented in Figure 10. In the initial model, 
there is a jump in $\eta$ at the boundary of the oxygen shell (corresponds to
the jump in composition). Due to mixing, this jump slightly moves towards lower
mass coordinate and becomes slightly wider. From comparison of Fig. 9 and 10 we 
see that the heating penetrates to deeper shells than composition changes.

\colfig{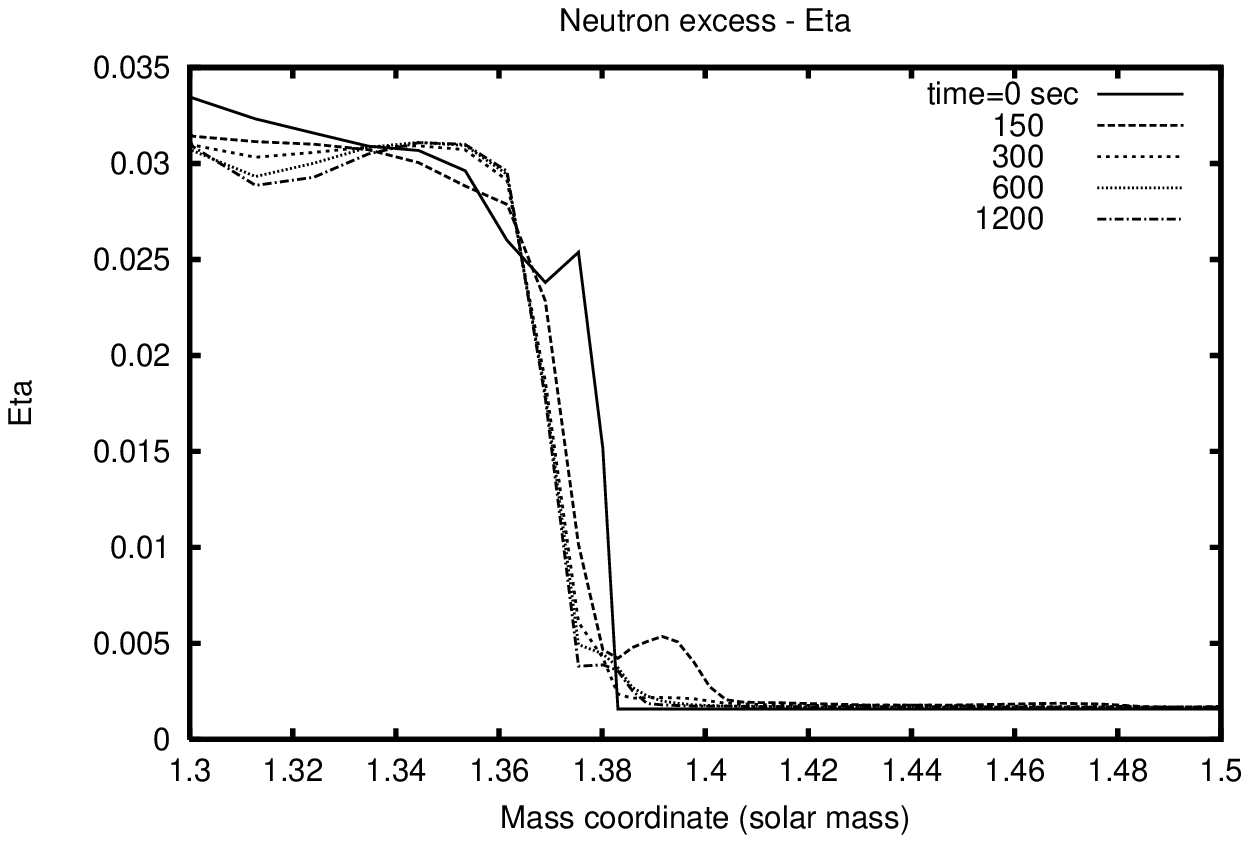}{8.0cm}{Neutron excess $\eta$ at the bottom}{f10}

There are two important consequences of this interaction:
(1) the heating and aditional oxygen and carbon in the few zones below the 1D 
oxygen convection shell cause these zones to be added to the oxygen shell
 and, together with few zones at the bottom of it, to be part of the burning 
zone (we can see a broader region with temperature above $2.2\times 10^9 \rm cm
$), (2) because of the cooling of the zones at the bottom of the oxygen 
shell, the temperature gradient becomes less than adiabatic in those zones and 
the convective flow exist as a result of a super adiabatic gradient at higher 
zones.

\colfig{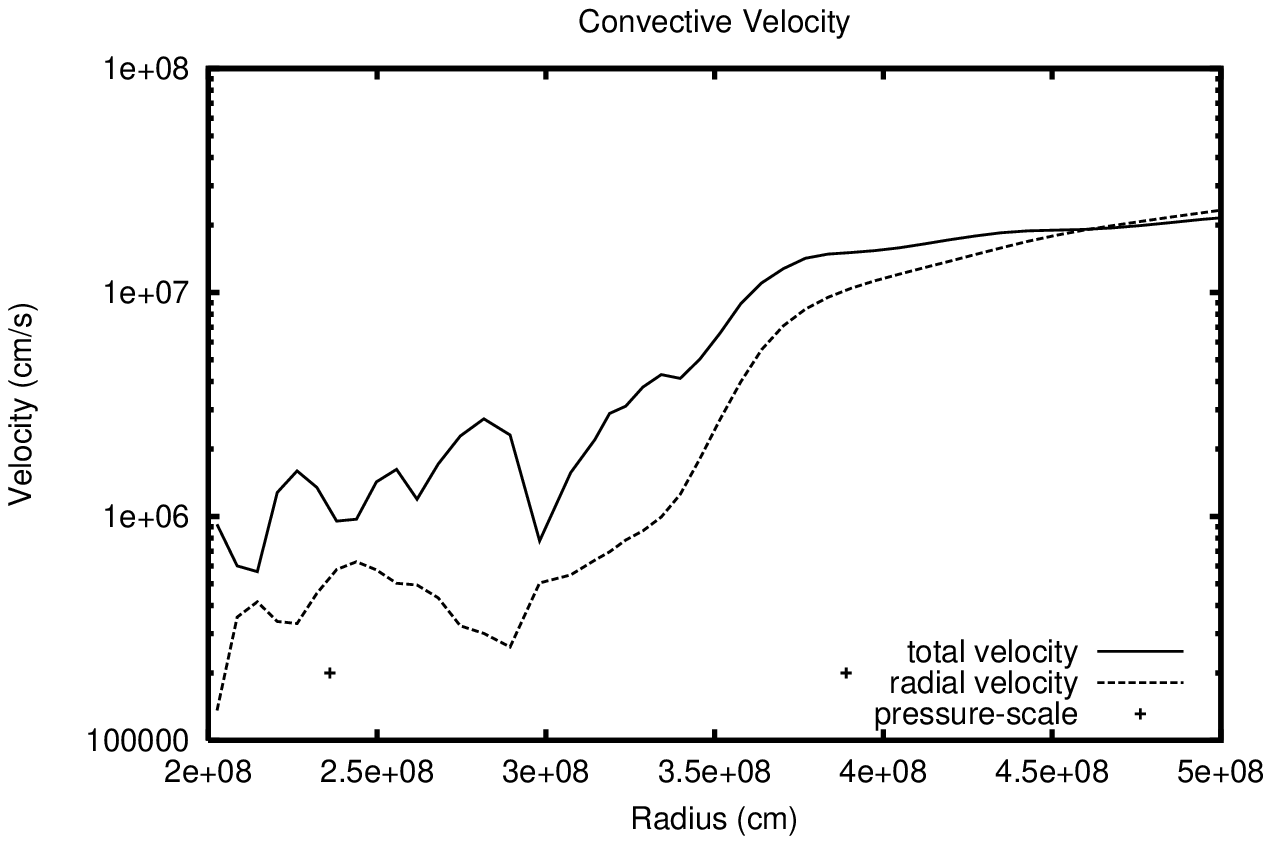}{8.0cm}{Convective velocity at the bottom}{f11}

Thus we see that the neighboring stable shells greatly affect the oxygen 
convective shell. The same convective flow that penetrates to the 
neighboring shells and allow the mixing of elements and energy between the 
shells affects further more the stable shells. In Figure 11 we present the
averaged convective velocity as a function of radius at the bottom of the 
oxygen shell. The dashed line corresponds to the fluctuations in radial 
component of the velocity, while the solid line corresponds to the amplitude of 
the fluctuating total velocity. The plus signs present points in which the 
pressure varies by a factor of ten. Three regions can be identified in this 
plot from right to left: a region of constant velocity, a region of exponential 
decrease, and another region of constant velocitis but with a noticeable 
difference between the total and the radial velocities. (i.e. the radial 
velocities are much smaller on average than the tangential velocities) which 
is indicative of g modes. This g modes can be identified in 2D presentation of 
the flow (Figure 12). 

\colfig{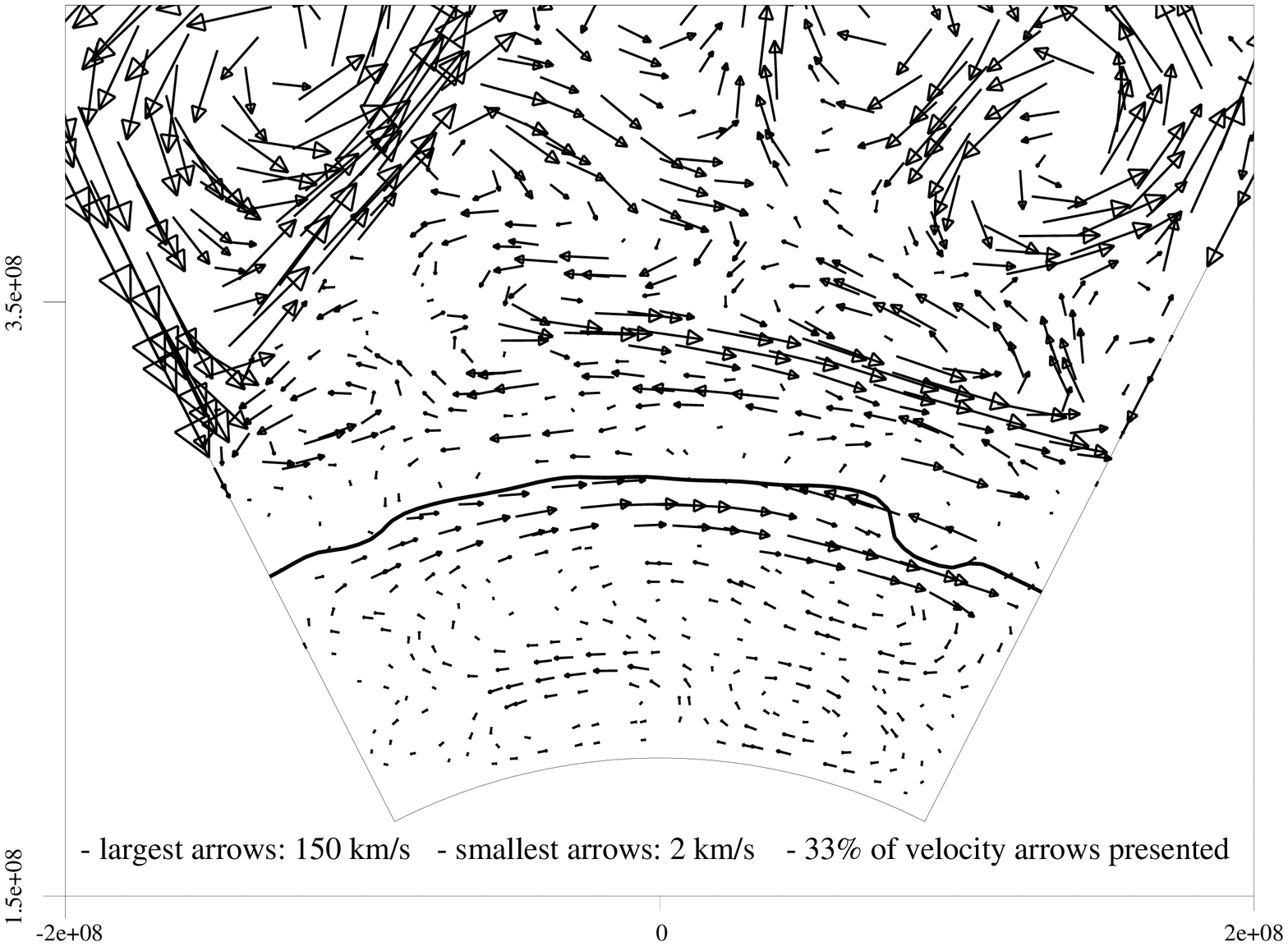}{8.0cm}{Velocity field at the bottom. The thick line represents
the composition jump that indicates the boundary of the oxygen shell}{f12}

In another simulation we included a much larger shell above the oxygen shell.
In this simulation we noticed similar flow characteristics of exponential
decay (Figure 13) and tangential preference of the flow (Figure 14). From Fig. 
11 and 13 we can also see that the exponential decrease at the bottom is on a
scale which is about one third of the pressure scale height, while at the top
it is over one pressure scale height.

\colfig{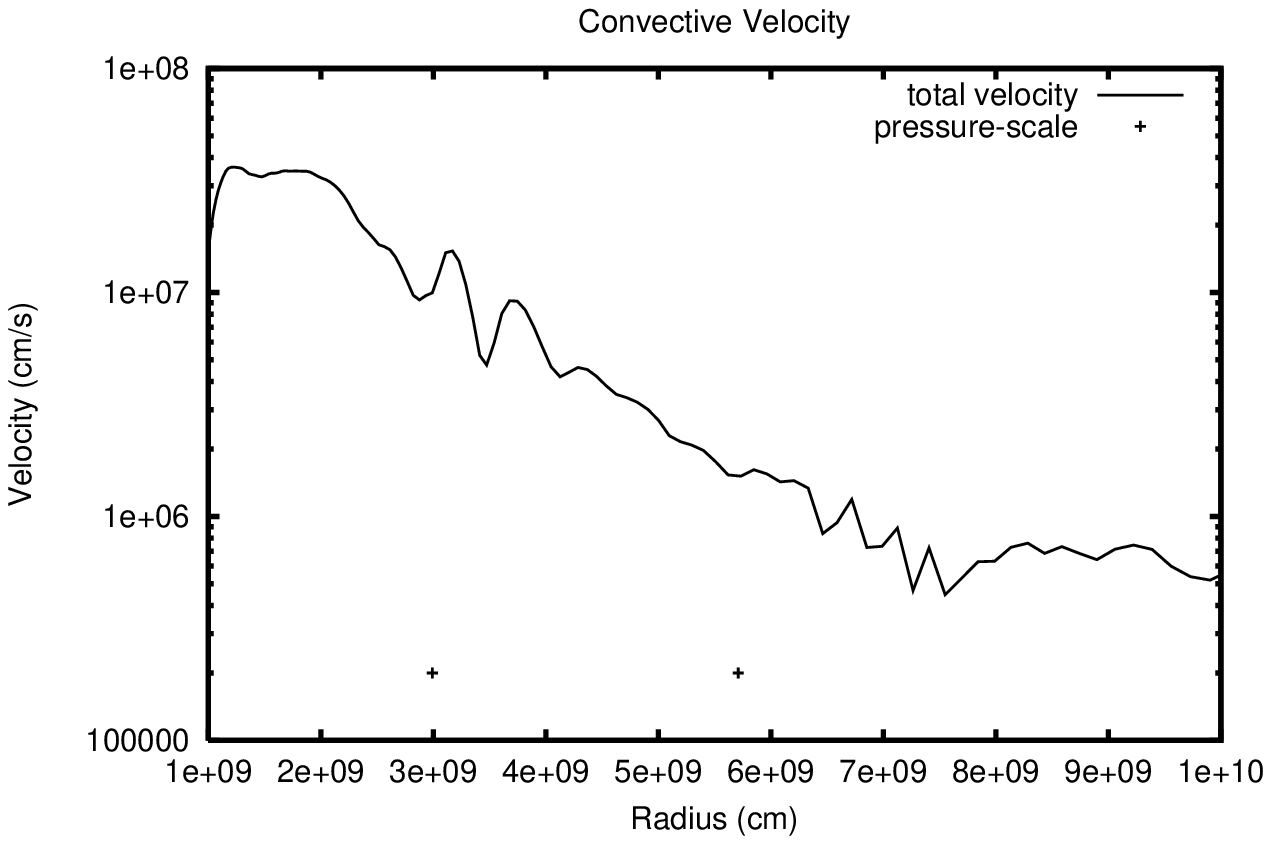}{8.0cm}{Convective velocity at the top}{f13}
\colfig{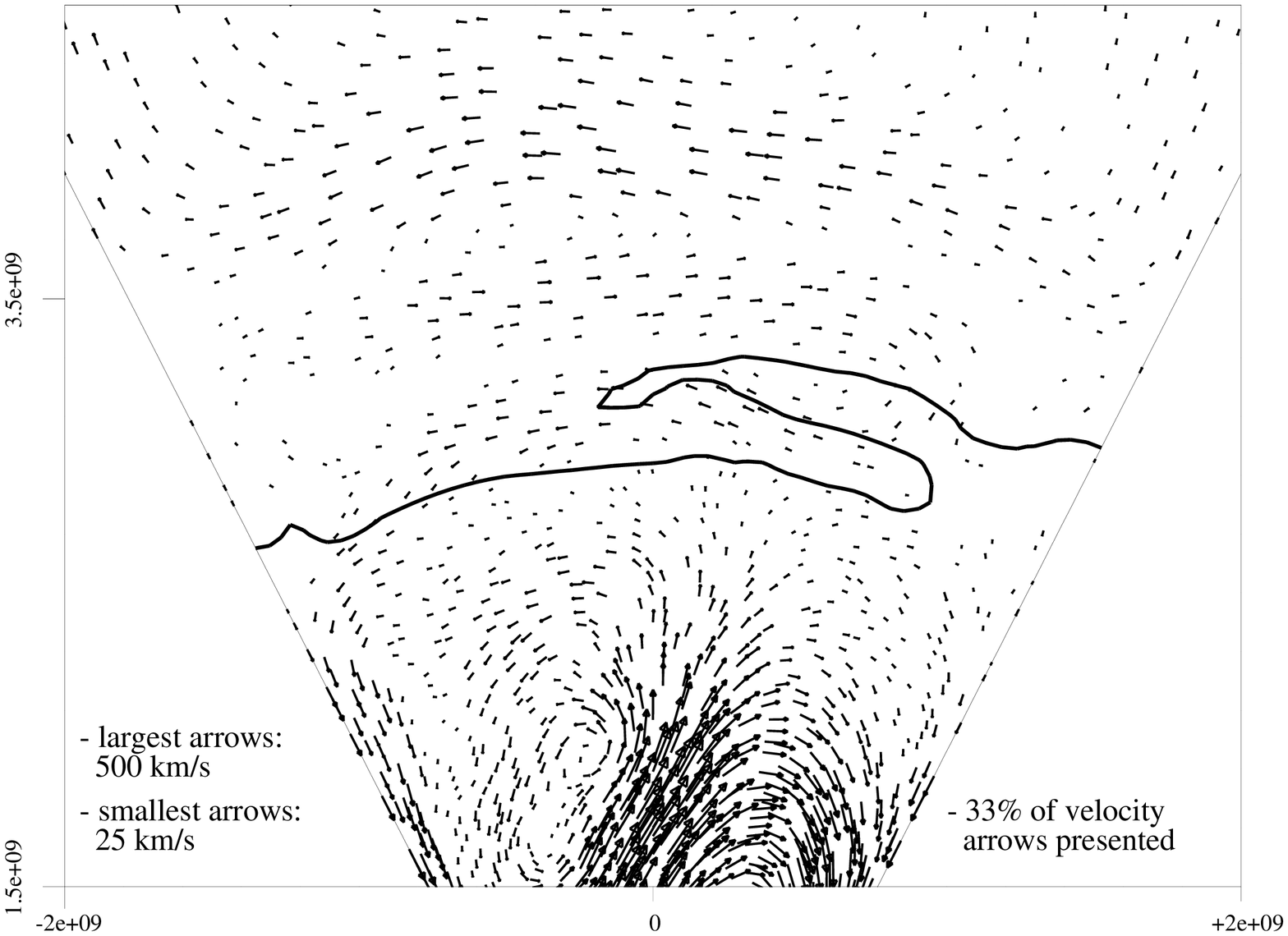}{8.0cm}{Velocity field at the top. The thick line represents
the composition jump that indicates the boundary of the oxygen shell}{f14}

\subsection{Numerical Sensitivity}

We performed several simulations with different parameters as mentioned in \S2. 
As our results are consistent with those of BA98, and since one of the main 
subjects in BA98 is the sensitivity of the results to numerical parameters, 
we would not present here all the results from our many simulations. Instead,
we will focus on the conclusions from those simulations: the most sensitive 
feature in the results
is the amount of heating below the oxygen shell. When there is more heating,
the temperature gradient is less than adiabatic in a larger portion of the 
(bottom of) oxygen shell. As a result, the convective flow is damped, carbon
can not reach the burning zone and we got less nuclear luminosity.

The opposite happened in simulations where we did not include the shell below
the oxygen shell: the temperature gradient was super adiabatic
so the velocities were higher (by a factor of five), more carbon entered the 
burning zone, and the nuclear luminosity was higher. From comparison of 
simulations with different resolution for this case (without the bottom shell),
 we found that the results are quite robust and not sensitive to the numerical 
resolution. 

In all of our simulations the other features were essentially the same, 
namely: the evolution of the flow from the bottom, the penetration to 
neighboring shells and the enrichment of the oxygen shell by carbon.

\colfig{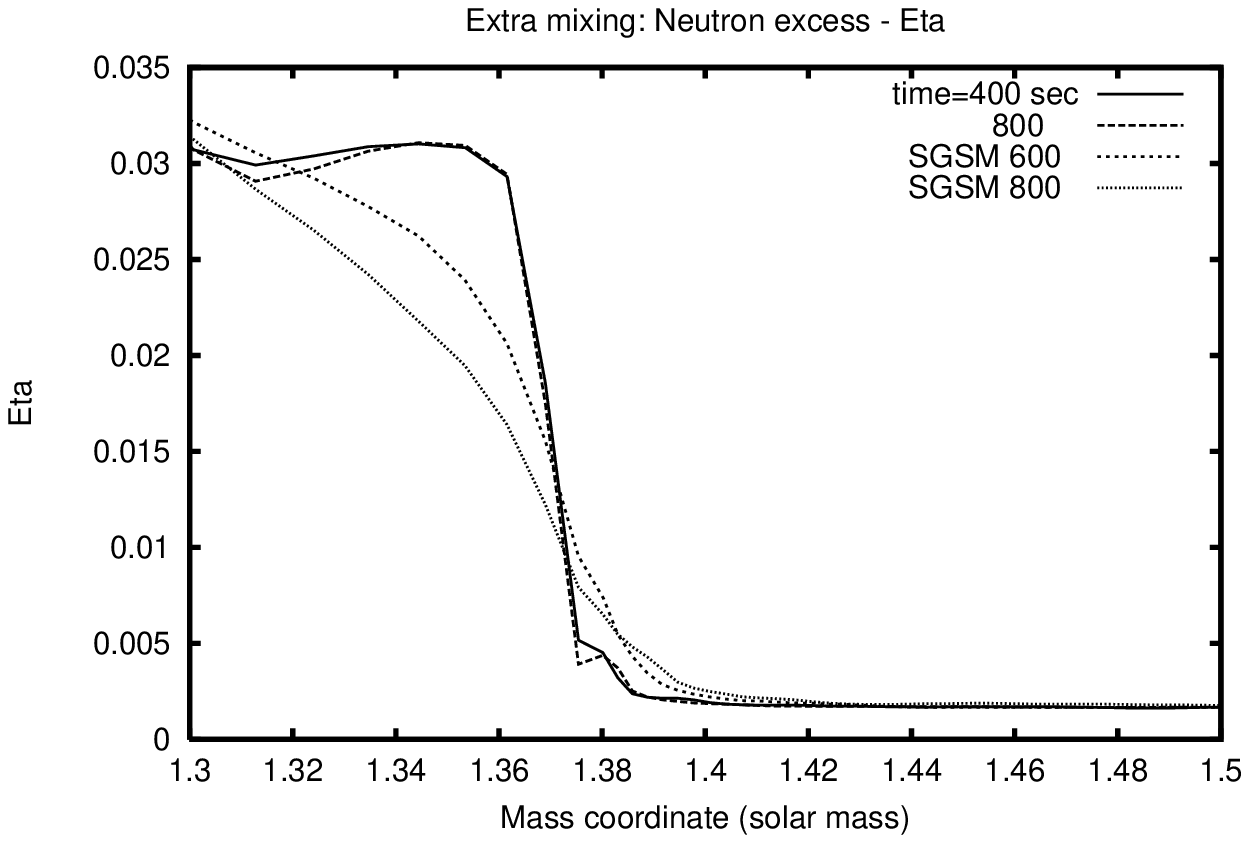}{8.0cm}{Extra $\eta$ mixing at the bottom}{f15}

In order to check extra mixing at scales shorter than numerical resolution,
we performed a simulation with SGSM mixing terms. This simulation started
from a 2D profile of the standard simulation at time 400~s and was carried
out to time 800~s. The differences between this simulation and the standard
simulation were: a more uniform composition at the oxygen shell - the RMS
fluctuations of carbon abundance at time 800~s were similar to the results
of the standard simulation at time 1200~s, and significantly more mixing at
the stable bottom shell. This can be seen in Figure 15 were we plot the
neutron excess $\eta$ for this simulation (SGSM) and the standard simulation. 
In this plot we can see a very small difference between the two profiles of
the standard simulation (at time 400~s - solid line and at time 800~s - long
dashed line). When SGSM terms were added, more mixing can be seen by time 600~s
(short dashed line), and additional mixing occurs until the end of the 
simulation (time 800~s - dotted line).

\section{Discussion}

When we compare our results to the results of BA98 we note that the evolution 
of the simulations is similar for times reached  previously. In both,
 convective flow evolves from bottom to top, with velocities exceeding $10\%$
 of the sound speed. The flow penetrates to the carbon rich region above the 
one dimensional edge of the convective region, and thus causes a significant 
enrichment of the convective region with carbon. These higher values of carbon 
abundances yield a much higher nuclear luminosity than did the original oxygen 
burning. 

However, at later times, we could identify signs of a steady state 
configuration in which the average abundances changed very slowly, and 
turbulent mixing within the convective region was able to decrease the 
fluctuations in the oxygen shell. However, density fluctuations near the 
boundaries of the convection zone (which are also composition discontinuities)
did {\em not} decrease at later times, and were equal to few percents.

The most prominent feature in our simulations is the interaction of the
convective shell with the neighboring shells. This interaction is due to 
penetration of the convective flow to the stable shells and the resulting mixing
of elements and energy. These effects were present in all of our simulations, 
and seemed to be physically resonable and valid. However, the exact amount of
mixing (especially at the bottom) depends on numerical parameters of the 
simulations. Moreover, since the initial 1D model did not include such overshoot,
we actually simulated a transient toward a more consistent model. For example,
we mentioned that the heating of the very few neighboring zones ,at the bottom
of the convection shell, lowers the temperature gradient below the adiabatic
value. If the convective flow is damped because of that, then the heat 
generation in these zones would evantually increase the temperature gradient, 
and convection would be restored.

The properties of the internal modes excited in the stable shells are undoubtedly
affected by the computational domain and the boundary condition. Reflective
waves are probably the cause for the constant level of velocities we noticed
at Figure 11. However, the volume below the oxygen shell in these stars is 
relatively small so a finite value of the velocity is quite plausible.

An interesting point is the validity of the results as we performed 2D 
simulations (and not 3D). This is not an easy question to answer, however, 
from various comparisons of 2D and 3D simulations (see for example 
\cite{krg99}) it seems that the main features of the interaction of the 
unstable layer and its neighboring layers are similar, and the overshoot and 
mixingin in both cases are comparable . As we do not claim to resolve this 
interaction quantitatively, we think that our results are valid.

Penetration of the convective flow to neighboring stable regions is a common 
feature that exists in many multidimensional simulations of convection, in a 
variety of stellar problems. \cite{hrl86} have studied such penetration beyond
a shallow convective region in 2D, and found that it generated internal g modes
in the stable region below the convection region, all the way to the bottom
boundary. Two regions in this stable layer can be identified from their figure
13: close to the unstable layer there is a sharp decrease in the kinetic energy,
but it does not go to zero, rather there is an almost constant level of kinetic
energy throughout the rest of the stable layer. \cite{krg99} have tested 
penetration in both 2D and 3D, they found that the energy fluxes in the stable 
layer in the 3D are about two thirds of the 2D results and that the constant
level is replaced with a continuas decrease when the viscosity coefficient is
increased at the bottom to avoid reflection of waves. 

In both these studies, ideal simplified physics was assumed. \cite{fls96}  
studied the structure and dynamics of shallow stellar surface convection zones,
using two dimensional radiation hydrodynamic simulations with more ``real'' 
physics. They found convective motions extending ``well beyond the 
boundary of convectively unstable region, with vertical velocities decaying 
exponentially with depth in the deeper parts of the lower overshoot region, 
as expected for linear $g^-$ modes.'' They suggested to approximate the average 
velocity at the stable region as the convective velocity at the boundary of the
unstable zone multiplied by an exponential decay factor with a lapse rate that 
is of order of the pressure scale height.

As explained by \cite{cox80} g$^-$ modes are unstable within convection
regions, and generally decrease exponentially with increasing distance from 
the boundaries of the oscillatory region. As \cite{fls96}
point out, \cite{ll59} have a relevant discussion in their
\S34, where they discuss some properties of potential flow. If compressibility 
and dissipation can be neglected, a steady state potential
flow which is periodic in some plane, must be damped exponentially in
a direction perpendicular to that plane. Also, the shortest wavelengths
will be damped fastest. Consequently, most of the overshoot
is given by the largest scales which are driven at the interface, that is,
the largest convective scale. 

We find a striking underlying unity in these results and ours, despite the 
significantly different astrophysical context involved: our simulations
correspond to deep shell burning convection that is driven by nuclear burning
at the bottom and cooled by neutrino radiation and in which spherical properties
of the domain are important; while these studies deal with envelope convection
driven by photon radiation at the surface in plane parallel geometry. In both
cases The flow extand beyond the formal boundaries of the convection zone, with
velocities that decay exponentially in the stable shells producing internal 
modes.

Near the boundaries of the convection zone, where the velocities are still 
large, quite efficient mixing is taking place. The amount of mixing beyond that
region, is not entirely clear. \cite{fls96} have used trajectories of test 
particles to estimate the diffusion coefficient and conclude that the diffusion 
coefficient corresponds directly to the exponentially decaying velocities.

Some studies of stellar evolution that used these mixing terms for hydrogen
core convection (\cite{hrw97}; Herwig \etal 1998), found that the lapse rate of 
decay of the diffusion should be much smaller ($\approx$2 percents) than the 
pressure scale height. They speculate that this is due to the large stability 
of the layers that are near the core convection zone. Yet another explanation
might be true, namely the flow characters at the stable shells and the many 
scales involved in mixing.

Mixing is a complex phenomenon which take place in a very small length scale 
comparing to the length scale of the star, and since stars have a lower 
effective viscosity than in the simulations, and since they evolve for longer 
times, an accurate prediction of mixing is very difficult. Further, we note 
that motion does not necessarily imply mixing! In particular, oscillatory 
potential flow, which is irrotational,would involve ``stretching'' and 
``contracting'' motions that do not give mixing, at least at lowest order.

As we can see in Figures 11-14, The flow in the stable shells is different 
than the flow inside the convection region, even though it is {\it not} purely 
potential flow (as revealed when examining the vorticity). Due to these 
differences in the flow, it seems plausible that the effective 
mixing velocity may be less than the hydrodynamic velocity, and its lapse rate 
different as well.  

As the various shells in our model have a different composition, mixing can be
directly examined from the 2D profiles and the 1D average of composition. 
However, as the scale of mixing is much smaller than numerical resolution, the 
results must be carefully examined. One way to examine the results is to compare
the mixing in the standard simulation to that in the simulation with sub grid 
mixing model. There are small differences in the average abundances in the 
oxygen shell between the two simulations, and the main difference is that with 
SGSM terms, the profile is more uniform because of the extra mixing. 
The difference between the simulations is much more prominent in the lower 
stable shell: the width of the abundance jump is much wider when SGSM terms were 
used, and penetrates to deeper zones, in which the ``constant level'' average 
velocity exists.

When we used SGSM we implicitly assumed that 
turbulent flow exists from the scale of the mesh resolution to the scale of 
molecular diffusion. Since this mixing occurs in a stable region, in which 
internal modes have been excited, the validity of this assumption is 
questionable. Thus it is fair to say that further work is needed to resolve 
this subject of mixing beyond the convective layers.

\cite{mdv99} have examined the effects of overshooting predicted by full 
spectrum turbulence model of convection (Canuto \& Mazzitelli 1991, 1992) 
within the context of lithium production in AGB stars. This is an interesting 
example of the way stellar convection may be tested.  A particular aspect of 
their work of interest here is the suggestion that {\it symmetric} overshoot 
may be constrained by existing observations. We note that a general feature of 
numerical simmulations is up-down asymmetry. Cooling flows are narrow, faster 
downdrafts than the corresponding updrafts, which are broader and slower. It is 
suggestive that in simulations of envelope convection driven by radiative 
cooling at the top, there is an overshoot at the bottom (\cite{hrl86,fls96}), 
while we, who have simulated shell convection driven by burning at the 
bottom, have found  exponential overshoot at the top (and the bottom). It may 
be that the causes of this asymmetry can be determined, so that a complete 
algorithm for the overshoot maybe devised for 1D stellar evolutionary 
calculations. 

In summary, we found that stellar evolution models should take into account 
penetration and an exponential decay of the velocities, and the excitation of 
internal modes, when simulating convective burning shells. Direct hydrodynamic 
simulation of stellar evolution places heavy demands on computer resources, 
especially for carbon and neon burning stages, which are slower. The tendency 
toward steady state revealed in our simulations is encouraging in that 
simplified algorithms may allow the construction of an improved set of one 
dimensional models.

We used an initial model which was derived from results of a one dimensional 
stellar evolution code, in which earlear stages of evolution were modeled with
standard MLT. Consequently, the profile we started with is unsatisfactory, and 
is not consistent with our conclusion that penetration and extra mixing should 
be taken into account. This should be kept in mind when using directly the 
results of our simulations.

\acknowledgments
Enlightening discussions with Grant Baz\`{a}n, Eli Livne and Falk Herwig, and
the interaction with the anonymous referee, are gratefully acknowledged.  This 
work was supported in part by DOE grant 
DE-FG03-98DP00214/A001.


\begin{thebibliography}{}
\bibitem[Arnett (1994)]{arn94}Arnett, D. 1994 \apj, 427, 932
\bibitem[Arnett (1996)]{arn96}Arnett, D. 1996, 
   {\it Supernovae and Nucleosynthesis\/} (Princeton, New Jersy: Princeton 
University Press)
\bibitem[Asida (2000)]{asd00}Asida, S.M. 2000 \apj, 528, 896
\bibitem[Asida \& Tuchman (1997)]{at97}Asida, S.M., \& Tuchman, Y. 1997 \apjl, 
491, L47
\bibitem[Baz\`{a}n \& Arnett (1994)]{ba94}Baz\`{a}n, G., \& Arnett, D. 1994, 
\apjl, 433, L41 
\bibitem[Baz\`{a}n \& Arnett (1998)]{ba98}Baz\`{a}n, G., \& Arnett, D. 1998, 
\apj, 496, 316
\bibitem[Canuto \& Mazzitelli 1991]{cm91}Canuto, V. M., \& Mazzitelli, I. 1991, 
\apj, 370, 295 
\bibitem[Canuto \& Mazzitelli 1992]{cm92}Canuto, V. M., \& Mazzitelli, I. 1992, 
\apj, 389, 724 
\bibitem[Cox (1980, \S 17.8)]{cox80}Cox, J.P. 1980, {\it Theory of Stellar 
Pulsation\/} (Princeton, New Jersy: Princeton University Press)
\bibitem[Deupree 1998]{dpr98}Deupree, R. G. 1998, \apj, 499, 340 
\bibitem[Freytag, Ludwig, \& Steffen (1996)]{fls96}Freytag, B., 
Ludwig, H. -G, \& Steffen, M. 1996, \aap, 313, 497 
\bibitem[Glasner \& Livne (1995)]{gl95}Glasner, S. A.,  \& 
Livne, E. 1995, \apjl, 445, L149 
\bibitem[Herwig \etal 1997]{hrw97}Herwig, F., Bl\"{o}ecker, T., 
Sch\"{o}enberner, D., \& El Eid, M. 1997, \aap, 324, L81 
\bibitem[Herwig, Sch\"{o}enberner, \& Bl\"{o}ecker 1998]{hrw98}Herwig, F., 
Sch\"{o}enberner, D., \& Bl\"{o}ecker, T. 1998, \aap, 340, L43 
\bibitem[Hurlburt, Toomre, \& Massaguer 1986]{hrl86}Hurlburt, N.E., Toomre, J., 
\& Massaguer, J.M. 1986, \apj, 311, 563
\bibitem[Kiraga \etal 1999]{krg99} Kiraga, M., Zahn, J.-P., 
St\c{e}pie\'{n}, K., Jahn, K., R\'{o}zyczka, M. \& Muthsam, H. J. 
1999, in ASP Conf. Ser. 173, Theory and Test of Convection in Stellar Structure,
ed. \'{A}.\ Gim\'{e}nez, E.\ Guinan, B.\ Montesinos (San Francisco: ASP), 269
\bibitem[Landau \& Lifshitz (1959)]{ll59} Landau, L. D., \& Lifshitz, E. M.,
    1959, {\it Fluid Mechanics} (London: Pergamon Press)
\bibitem[Livne 1993]{lvn93}Livne, E.  1993, \apj, 412, 634
\bibitem[Mazzitelli, D'Antona, \& Ventura (1999)]{mdv99}
Mazzitelli, I., D'Antona, F., \& Ventura, P. 1999, \aap, 348, 846 

\bibitem[Smagorinsky (1963)]{smg63}Smagorinsky, J. 1963, Mon. Weather Rev., 
91(3) 99

\end{thebibliography}
\end{document}